# Genetic Algorithm with Ensemble Learning for Detecting Community Structure in Complex Networks

Dongxiao He
College of Computer
Science and Technology
Jilin University
Changchun, China
hedongxiaojlu@gmail.com

Zhe Wang
College of Computer
Science and Technology
Jilin University
Changchun, China
wz2000@jlu.edu.cn

Bin Yang
College of Computer
Science and Technology
Jilin University
Changchun, China
yangbin@jlu.edu.cn

Chunguang Zhou
College of Computer
Science and Technology
Jilin University
Changchun, China
cgzhou@jlu.edu.cn

*Abstract*—Community detection in complex networks is a topic of considerable recent interest within the scientific community. For dealing with the problem that genetic algorithm are hardly applied to community detection, we propose a genetic algorithm with ensemble learning (GAEL) for detecting community structure in complex networks. GAEL replaces its traditional crossover operator with a multi-individual crossover operator based on ensemble learning. Therefore, GAEL can avoid the problems that are brought by traditional crossover operator which is only able to mix string blocks of different individuals, but not able to recombine clustering contexts of different individuals into new better ones. In addition, the local search strategy, which makes mutated node be placed into the community where most of its neighbors are, is used in mutation operator. At last, a Markov random walk based method is used to initialize population in this paper, and it can provide us a population of accurate and diverse clustering solutions. Those diverse and accurate individuals are suitable for ensemble learning based multi-individual crossover operator. The proposed GAEL is tested on both computer-generated and real-world networks, and compared with current representative algorithms for community detection in complex networks. Experimental results demonstrate that GAEL is highly effective at discovering community structure.

I. INTRODUCTION

In the real world, many complex systems take the form of networks. Examples include biological networks such as metabolic networks and neural networks, technological networks such as the Internet and the World-Wide Web, and social networks such as scientific collaboration networks and email networks. Networks, in general, are constituted by a set of nodes or vertices and a set of links or edges, where a node is an individual member in the complex system and an edge is a link between nodes according to a relation in the system [1]. The study of complex networks revealed features as small-world character and scale-free degree distribution [2, 3].

Recently, the characterization of community (or module) structures [4] in complex networks has been paid a considerable amount of attention by the scientific community [4-9]. A community in a network is usually thought of as a group of nodes that are similar to each other and dissimilar from the rest of the network. It's believed that nodes are densely connected within communities while being loosely connected to the rest of the networks [4]. As community structure is similar to cluster in data clustering, the property of community structure is also sometimes called clustering. Such communities have been observed in many different contexts, including biological networks, technological networks, social networks, etc. Detecting the partitioning of a network in clusters could clearly have important theoretical significance and practical value to analyze the topology structures of complex networks, understand the functions of complex networks, and predict the behaviors of complex networks [4][5].

As the identification of communities has important theoretical significance and practical value, the problem of community detection has been receiving a lot of attention and many different algorithms have been proposed [4-13]. These algorithms can be divided into two categories: heuristic and optimization based methods. The former solves the problem of community detection based on some intuitive assumptions or heuristic rules. For example, the heuristic rule used in Girvan-Newman (GN) algorithm [4] is that the "edge betweenness" of inter-community links should be larger than that of intra-community links. The latter solves the problem of community detection by transforming it into an optimization problem and trying to find an optimal solution for a predefined objective function such as the modularity $Q$ employed in several algorithms [10-13]. Unfortunately, maximizing the modularity $Q$ has been proven to be a nondeterministic polynomial time (NP)-complete problem [8], which makes it unable to carry out an exhaustive search of all possible divisions for the optimal value of $Q$ in a large network. Genetic algorithm (GA) is an efficacious method for solving NP-complete problems, and able to dramatically reduce the time complexity for solving the problems while ensuring the quality of the solutions. The existing GA based algorithms for community detection are not exempt from two drawbacks: slow convergence and low precision. The major cause which leads to these drawbacks is that commonly used crossover operators can not be suitable for community detection in complex networks. The existing genetic based algorithms for community detection remove the crossover operator from genetic algorithm [12] or can't provide an effective crossover operator [11] [13]. Consequently, the search capability of GA is weakened. In addition, generating initial population randomly and changing the value of a gene at random in mutation also make genetic based algorithms for community detection ineffective.

In this paper, a genetic algorithm with ensemble learning (GAEL) for community detection in complex networks, which is able to avoid the defects of existing genetic based algorithms, is presented. GAEL replaces its traditional crossover operator with an ensemble learning [14] based multi-individual crossover operator. Therefore, GAEL can avoid the problems that are brought by traditional crossover operator which ignores the clustering contexts and only exchanges string blocks of different individuals. Furthermore, ensemble learning based multi-individual crossover operator can transmit the excellent characteristics from one generation to the next, and the global search ability of crossover operator can be brought into full play. Local search strategy, which makes mutated node be placed into the community where most of its neighbors are [7], is used in mutation operator. Compared with commonly used mutation operators, specialized mutation operator can remove useless iterations and reduce the research space of possible solutions thus speeding up the convergence of GAEL. With reference to [15], a Markov random walk based population initialization method is proposed in this paper, and it can provide us a population of accurate and diverse clustering solutions. Those diverse and accurate individuals are suitable for ensemble learning thus ensemble learning based multi-individual crossover operator can achieve a better result. Furthermore, to some extent the space where initial population produced by this method is approaches the space where optimal solution is thus speeding up the search of GAEL.

## II. GENETIC ALGORITHM WITH ENSEMBLE LEARING FOR COMMUNITY DETECTION IN COMPLEX NETWORKS

### A. Problem Definition and Individuals Encoding

A complex network can be defined as a graph $G=(V, E)$, where $V$ is the set of vertices, and $E$ is the set of edges connecting pairs of vertices. Detecting community structure is to find the division of network nodes into groups within which the network connections are dense, but between which they are sparser. In order to quantify the strength of a particular division of the network, Newman and Girvan proposed a quantitative measure called modularity $Q$ [9]. This quantity measures the fraction of the edges in the network that connect vertices of the same type minus the expected value of the same quantity in a network with the same community divisions but random connections between the vertices, and it can be calculated as:

$$Q = \sum_i (e_{ii} - a_i^2), \quad (1)$$

where $i$ is the index of the communities, $e_{ii}$ is the fraction of edges that connect vertices in the community $i$ and $a_i$ is the fraction of edges with at least one node in the community $i$.

Modularity $Q$ suffers from resolution limit problem as shown by Fortunato and Barthelemy [16], where small modules tend to merge into bigger one. However, modularity $Q$ is still an effective quality metric for assessment of partitioning a network into communities at present, and has been widely used in recent studies [8][10-13]. Therefore, GAEL also employs it as objective function, and this objective function is to be maximized.

In this paper, we adopt the string-of-group encoding strategy. Each candidate clustering solution is coded as an integer string and the value of an integer in the string represents the identifier of the community in which the node is classified.

### B. Population Initialization

Theoretical and experimental results clearly showed that ensemble learning is effective only when the individual solutions are accurate and diverse [17]. In order to satisfy this condition and give full play to the global search ability of the ensemble learning based multi-individual crossover operator, we proposed the Markov random walk based population initialization method referring to [15].

We suppose that an agent freely walks from one vertex to another along the links between them. When the agent arrives at a node, it will select one of its neighbors at random and go there. Let $p_{ij}$ be the probability of the agent walking from node $i$ to its neighbor node $j$ through one step walking, then it can be computed as (here we only consider unweighted and undirected networks)

$$p_{ij} = \frac{1}{d_i}, \quad (2)$$

where $d_i$ is the degree of node $i$.

Let $P_t^l(i)$ be the probability that the agent starting from node $i$ can eventually arrive at a specific destination node $t$ within $l$ steps, its value can be estimated iteratively by

$$P_t^l(i) = \begin{cases} 1 & if \ i=t \\ \sum_{\langle i,j \rangle} p_{ij} \cdot P_t^{l-1}(j) & if \ i \neq t \end{cases} (1 \leq i,t \leq n), \quad (3)$$

where $<i, j>$ denotes the link connecting nodes $i$ and $j$, and $n$ is the number of nodes in the network.

The agent's walk can be viewed as a stochastic process defined based on the links' attributes. As the link density within a community is, in general, much higher than that between communities, agents that start from nodes within the community of a destination node should have more paths to choose from in order to reach the destination node within some $l$ steps, where the value of $l$ cannot be too large. On the contrary, agents that start from nodes outside the community of the destination node have a much lower probability of eventually arriving at the destination.

Using the above theory, the algorithm for Individual Generation based on Markov Random Walk (IGMRW) is proposed which will be described in the following part, and it's a recursive one.

Step 1    Randomly select the destination node $t$ from the network (or sub-network), and calculate $P_t^l(i)$ for each node $i$.

Step 2 Rank all the nodes according to their associated probability values, and find a cutoff point along which the split can result in the greatest increase in global modularity.

Step 3 If there is no cutoff point that can result in an increase in global modularity, the recursive procedure is terminated. Else, divide the network (or sub-network) into two parts according to the cutoff point found by Step 2, and iterate the procedure (going back to Step 1) for all the sub-networks.

In each recursive call procedure, randomly select the destination node $t$ from the sub-network in Step 1, and so different solutions (individuals) can be obtained by applying IGMRW multiple times on the same network. Therefore, IGMRW can provide diverse individuals. In addition, individuals generated by the combined action of Markov random walk theory and network modularity $Q$ have a certain precision.

Therefore, it can be seen that IGMRW can produce diverse and accurate individuals, and these individuals are very suitable for ensemble learning. Furthermore, to some extent, the space where initial individuals produced by this method are approaches the space where optimal solution is, as a result, it can speed up the convergence of GAEL.

## C. Crossover Operator

Crossover operator is the key operator of genetic algorithm, and plays an important part in global searching. According to special requirements of genetic based algorithms for community detection for crossover operator, an ensemble learning based multi-individual crossover operator is proposed in this chapter.

As community structure is a relational property, the label of community is only an identifier. The same label in different individuals may correspond to different communities, even in different individuals with the same clustering context the same label may correspond to different communities. For example, both the individual (1,1,1,2,2,2) and the individual (2,2,2,1,1,1) represent the same clustering solution, but the community whose label is 1 in the first individual corresponds to the community whose label is 2 in the second individual. After executing the one-point crossover operator and selecting 3 as crossover site, their offspring (1,1,1,1,1,1) and (2,2,2,2,2,2) are significantly different from their parents and have no meaning. This above example lets us know that the traditional crossover operator of GA such as the one-point crossover operator is only able to mix string blocks of different individuals, but not able to recombine clustering contexts of different individuals into new better ones. The offspring which don't inherit the excellent characteristics from their parents is called adverse individual in this paper. The traditional crossover operator often produces adverse individuals and leads to the disruption of good building blocks, thus significantly degrades the search capability of GA.

A large number of ensemble learning methods for data clustering have existed, and generally they work with combining multiple clustering results into a single consensus one by leveraging their consensus [14]. It is noted that the function of the ensemble learning method for combining multiple clustering solutions into a single consensus one is somewhat similar to that of a crossover operator of GA that works to mix different candidate clustering solutions into a new better one [18]. In this paper, traditional crossover operator with two individuals is extended to multi-individual crossover operator, and then an ensemble learning based multi-individual crossover operator is proposed for GAEL.

Before describing our proposed crossover operator in detail, we will introduce two relevant concepts next.

DEFINITION 1 (edge join strength). It is assumed that $M$ community divisions are provided. The edge join strength $c_{vw}$ of link $<v, w>$ in the network can be calculated as

$$c_{vw} = \frac{k_{vw}}{M}. \quad (4)$$

where $k_{vw}$ denotes the number of times the link $<v, w>$ is considered as intra-community connection among the $M$ partitions.

On the basis of several given community divisions, edge join strength is given by ensemble learning method, and it is a quantity that can single out edges connecting nodes belonging to the same community. In other words, $c_{vw}$ can be considered as the probability that link $<v, w>$ is an intra-community connection. Obviously, the larger the value of $c_{vw}$ is, the greater the tendency link $<v, w>$ has to be an intra-community connection. On the contrary, link $<v, w>$ is more likely to be an inter-community connection. Therefore, it is considered the edge join strength of intra-community connection should be larger than that of inter-community connection if the community structure of network is reasonable.

DEFINITION 2 (edge structural similarity). A network can be defined as a graph $G = \{V, E\}$, where $V$ is the set of nodes, and $E$ is the set of edges. For an edge connecting nodes $v$ and $w$ in the network, its edge structural similarity $\sigma(<v, w>)$ can be computed as

$$\sigma(<v, w>) = \frac{|\Gamma(v) \cap \Gamma(w)|}{\sqrt{|\Gamma(v)||\Gamma(w)|}}. \quad (5)$$

where $\Gamma(v) = \{w \in V \mid <v, w> \in E\} \cup \{v\}$, and $\Gamma(v)$ denotes the neighborhood of node $v$.

We generalize vertex structural similarity in [6] to edges, and edge structural similarity is defined by using network structure. This similarity metrics is built on the basis of the acquaintance model in sociology, and its key idea is as follows: the more friends two persons share, the more familiar they may be with each other. Therefore, the greater the similarity between the two nodes, the larger the value of edge structural similarity, in other words, this edge is more likely to be an intra-community connection.

Using the above theory, ensemble learning based multi-individual crossover operator can be described as follows:

Step 1 Select $M$ ($M<N$) promising clustering solutions from the present population by tournament selection;
// *Selection pressure can be easily adjusted by changing the tournament size.*

Step 2    Sort all the links in descending order according to their associated edge join strength values;
Step 3    Sort the links having identical edge join strength in descending order according to their edge structural similarity values;
Step 4    Set the sequence of links arranged by step 2 and 3 to *X*;
Step 5    Generate a new individual (that is a new community division) using agglomerative hierarchical clustering method;
  Step 5.1    Set the initial state: each vertex is the sole member of one of *n* communities;
  Step 5.2    Get the first link in *X*, and delete it from *X*;
  Step 5.3    If the link got by Step 5.2 connects two different communities, join the communities together and recalculate the value of *Q* for the new community division;
  Step 5.4    If *X* is not null, go to Step 5.2;
  Step 5.5    Select the community division with the maximal value of *Q* from *n* different community divisions produced by the above process, and take it as the new individual generated by the multi-individual crossover operator based on ensemble learning. ∎

Edge join strength gives a measurement to judge whether the edge is an intra-community edge by utilizing the information of several given community divisions and ensemble learning, while edge structural similarity gives the same measurement from other view by utilizing the local information of network topology structure. As the concept of edge join strength utilizes rationally prior knowledge, it is a more accurate measurement. Therefore, edge join strength is used as primary measure standard while edge structural similarity is used as auxiliary measure standard which is used only when edges have identical value of edge join strength.

In this paper, the information of given community divisions is transformed into edge join strength by ensemble learning method. This conversion utilizes a measure that is "whether the edge is an intra-community edge" rather than community identifiers. This is because this measure is in one-to-one correspondence with the information of community divisions but community identifiers are not. For example, the individual *A*=(1,1,1,2,2,2) and the individual *B*=(2,2,2,1,1,1) represent the same community division where nodes $\{v_1,v_2,v_3\}$ are classified into one community and nodes $\{v_4,v_5,v_6\}$ are classified into other community. The community identifiers of both nodes $v_2$ and $v_3$ are 1 in individual *A*, while they are 2 in individual *B*. However, the link $\langle v_2, v_3 \rangle$ is an intra-community connection whether in individual *A* or in individual *B*. And here it becomes evident, that our proposed method can avoid the problem caused by traditional crossover operator which only exchanges string blocks of different individuals without consideration of the clustering contexts. Moreover, multi-individual crossover operator based on ensemble learning can utilize rationally the efficient clustering information of parents and generate better offspring, thus global searching ability of crossover operator is strengthened.

## D. Mutation and Selection Operator

Mutation operator is embedded into genetic algorithm to reinforce its ability of local search. We adopt the local search strategy [7] to realize mutation. We perform mutation to some number of randomly selected genes. In mutation function, a node adopts the label that most of its neighbors currently have, and we break ties randomly among the possible candidates when the node has an equal maximum number of neighbors in two or more communities. Comparing with traditional mutation operators that randomly change the label of a node, specialized mutation operator in this paper can decrease useless exploration and allows to reduce the research space of possible solutions with pertinence thus speeding up the search of GAEL. Furthermore, specialized mutation operator is able to effectively solve the problem, that is, there may be a small number of misplaced nodes that do not affect the overall fitness value very much.

Selection operator plays a role in global searching. In order to retain the fittest individual in each generation and improve the convergence speed of genetic algorithm, μ+λ selection for which genetic algorithms for combinatorial optimization have a partiality is used in this paper. The process of μ+λ selection can be described as follows: let the size of parent population be μ, generate λ offspring from randomly chosen parents, and single out μ best among parents and offspring as next population.

In addition, GAEL is terminated on condition that at least one of the following two requirements is met: 1) The fittest individual in the population has not been improved for a predefined number of generations; 2) The number of the current generation is beyond limitation.

## III. EXPERIMENTAL RESULTS

In order to quantitatively analyze the performance of GAEL, we have applied it to computer-generated and real-world networks whose community structure is already known.

## A. Computer-generated networks

In this section we tested the performance of GAEL on computer-generated networks with known community structure. This experimental method has been widely used and considered as benchmark [10]. We have generated using a computer a series of artificial networks. Each network consists of *n* = 128 nodes divided into four communities of 32. Each node has on average $k_{in}$ edges connecting it to members of the same community and $k_{out}$ edges to members of other communities, with $k_{in}$ and $k_{out}$ chosen such that the total expected degree $k_{in} + k_{out} = 16$, in this case. Let $P_{out}$ denotes the fraction of edges that fall within communities to the total number of edges in the network. As $P_{out}$ is increased from zero, community structures become more diffused and the resulting networks pose greater and greater challenges to the community-finding algorithm. We generated 100 different networks for values of $P_{out}$ ranging from 0 to 0.5 to test our algorithm.

We calculated the fraction of vertices correctly classified and took it as a measure of clustering accuracy. This measurement was used first by Newman [4], and has been the most commonly used method of clustering accuracy. In order to investigate the performance of GAEL, the accuracy of our algorithm is compared with the Girvan-Newman algorithm (GN) [4] and Newman-fast algorithm (FN) [10] who are quite classical and frequently referenced at present, and the results are reported in Fig. 1(a). For the sake of further verifying the effectiveness of GAEL, the accuracy of GAEL is also compared with other two excellent algorithms at present, the CPM algorithm [5] and FEC algorithm [15], and the results are reported in Fig. 1(b). In Fig.1 we show the fraction of vertices correctly assigned to the four communities by the algorithm, averaged over the 100 runs, when $P_{out}$ increases from 0 to 0.5. As shown in Fig. 1(a), our algorithm significantly outperforms GN and FN, moreover, as $P_{out}$ becomes larger and larger, the superiority of our algorithm becomes more and more significant. For example, when $P_{out}$ equals 0.4 our algorithm performs perfectly, classifying virtually 100% of vertices into their correct communities, while GN correctly classified 26.5626% of vertices and FN correctly identified 84.76565%; when $P_{out}$ increases to 0.5 at which the number of within-community and between-community edges per vertex is the same GAEL correctly identifies an average of 99.2188% of vertices, while GN and FN correctly identified 26.5625% and 67.6563% respectively which are obviously lower than the accuracy of GAEL. As shown in Fig. 1(b), GAEL significantly outperforms CPM and FEC. As $P_{out}$ becomes large and large, the superiority of our algorithm over CPM and FEC becomes more and more significant. For example, when $P_{out}$ equals 0.45 our algorithm can maintains the clustering accuracy at 99.2188%, while FEC reduces the clustering accuracy to 90.625% and CPM reduces the clustering accuracy to 61.7188%.

### B. Real-world network

Real-world networks usually have different topological properties from the computer-generated networks, and so we further test our algorithm on data from real-world networks. To this end, we have selected two datasets representing real-work networks which have been widely used as a test case for new methods for complex networks. Since GAEL takes modularity $Q$ as the object function, we compare our algorithm with two algorithms optimizing $Q$ function which are selected from the previous section, the GN algorithm and FN algorithm.

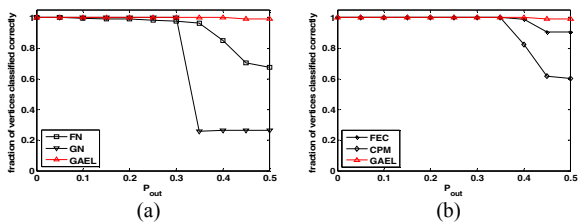

Figure 1. The fraction of vertices correctly classified by different algorithms as Pout is varied: (a) Comparison of GAEL, GN and FN algorithms; (b) Comparison of GAEL, CPM and FEC algorithms.

*1) Zachary's karate club network:* Zachary's karate club network [19] is drawn from the well-known "karate club" study of Zachary. It consists of 34 members of a club at an American university as nodes and 78 edges representing friendships among members of the club, which was recorded over a two-year period by Zachary. During the course of his study, a dispute between the club's administrator and its principle karate teacher arose and the club eventually split into two factions, centered on the administrator and the teacher.

Applying the GAEL algorithm ten different times on this network, we obtained ten identical community divisions. We show the community structure found by our algorithm in Fig. 2(a). The principle karate teacher and the club's administrator are represented by nodes 1 and 34 respectively in Fig. 2(a). Circles represent members associated with the club administrator's faction while squares represent members associated with the principle karate teacher's faction. GAEL divides this network into four groups as indicated with four different colors. From Fig. 2(a) we can see, not only are the two factions well separated according to the reality but also GAEL divides them further into four smaller groups, and the two groups separated by the red line are consistent with disruption among members in reality. The modularity $Q$ obtained by GAEL is 0.4198 which is higher than the modularity $Q = 0.3715$ for the actual division of the club members following the break-up. Compared with other algorithms, the modularity $Q=0.4198$ obtained by GAEL is greater than the value of $Q=0.4013$ arrived at with GN and the value $Q=0.381$ with FN.

*2) American college football network:* As a further test of our algorithm, we apply it to the American college football network which is constructed based on the schedule of Division I games during the 2000 season [4]. This network contains 115 nodes and 616 edges, nodes represent college football teams and edges represent games between the two teams they connect. These teams are divided into "conferences", with intra-conference games being more frequent than inter-conference games. The real community structure is the conferences that each team belongs to. Ten identical community divisions are obtained from applying the GAEL algorithm ten different times on this network, and the community division is depicted in Fig. 2(b). The different shapes combined with colors represent 12 conferences in reality in Fig. 2(b). As shown in Fig. 2(b), the GAEL algorithm reveals ten communities, and most teams are correctly grouped with the other teams in their conference; there are a few independent teams that do not belong to any conference, and these teams are grouped with the conference with which they are most closely associated by our algorithm. The result obtained by GAEL is compared with those obtained by GN and FN, and the comparison is listed in TABLEⅠ. As shown in TABLEⅠ, both the modularity Q and the clustering accuracy obtained by GAEL are obviously higher than those obtained by GN and FN.

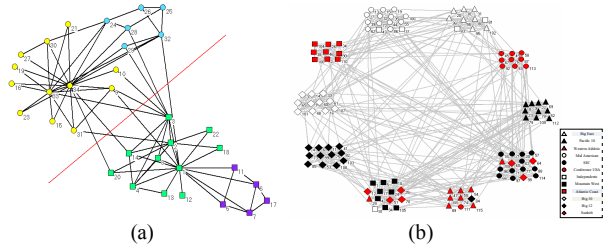

Figure 2. Testing GAEL on real-word networks: (a) Community structure identified by GAEL on Zachary's karate club network; (b) Community structure identified by GAEL on American College Football network.

TABLE I. COMPARISON OF SOLUTIONS OF DIFFERENT ALGORITHMS

| Average value | GN | FN | GAEL |
|---|---|---|---|
| Modularity $Q$ | 0.6005 | 0.546 | 0.6054 |
| Clustering accuracy | 82.61% | 62.61% | 86.09% |

## IV. CONCLUSIONS

In this paper, a genetic algorithm with ensemble learning for community detection in complex networks has been proposed. Experimental results on both computer-generated and real-world networks have demonstrated its effectiveness. The characteristics of GAEL are concluded as follows: first, the population of GAEL is initialized by using the Markov random walk based population initialization method which can produce accurate and diverse individuals that are suitable for ensemble learning. Second, GAEL replaces traditional crossover operator of genetic algorithm with an ensemble learning based multi-individual crossover operator which can improve the global search ability of GAEL. Lastly, the local search strategy is introduced into mutation operator thus enhancing the local search performance of GAEL. In the future we plan to raise the efficiency of crossover operator of GAEL by improving the method of ensemble learning and apply GAEL to analyze biological networks such as metabolic networks, etc.

ACKNOWLEDGMENT

This work was supported by (1) the National Natural Science Foundation of China under Grant No. 60673099 and No. 60873146, (2) National High Technology Research and Development Program of China under Grant No. 2007AAO4Z114 and No. 2009AA02Z307, (3) the Key Laboratory for Symbol Computation and Knowledge Engineering of the National Education Ministry of China.

REFERENCES

[1] M. A. Porter, J. P. Onnela and P. J. Mucha. "Communities in Networks". arXiv:0902.3788v1 [physics.soc-ph], 2009.
[2] S. Milgram. "The small word problem". Psychology Today, vol. 2, Jan. 1967, pp. 60-67.
[3] R. Albert, H. Jeong and A. L. Barabási. "The Internet's Achilles heel: Error and attack tolerance of complex networks". Nature, vol. 406, Jul. 1967, pp. 378-382.
[4] M. Girvan and M. E. J. Newman. "Community structure in social and biological networks". Proc. Natl. Acad. Sci. USA, vol. 99, Dec. 2002, pp. 7821-7826, DOI: 10.1073/pnas.122653799.
[5] G. Palla, I. Derényi, I. Farkas and T. Vicsek. "Uncovering the overlapping community structure of complex networks in nature and society". Nature, vol. 435, Jun. 2005, pp. 814-818, DOI: 10.1038/nature03607.
[6] X. Xu, N. Yuruk, Z. Feng and T. A. J. Schweiger. "SCAN: A Structural Clustering Algorithm for Networks". Proceedings of the 13th ACM SIGKDD International Conference on Knowledge Discovery and Data Mining (KDD 07), ACM Press, Aug. 2007, pp. 824-833, DOI: 10.1145/1281192.1281280.
[7] U. N. Raghavan, R. Albert and S. Kumara. "Near linear time algorithm to detect community structures in large-scale networks". Physical Review E, vol. 76, Sept. 2007, pp. 036106, DOI: 10.1103/PhysRevE.76.036106.
[8] Y. Hu, M. Li, P. Zhang, Y. Fan and Z. Di. "Community detection by signaling on complex networks". Physical Review E, vol. 78, Jul. 2008, pp. 016115, DOI: 10.1103/PhysRevE.78.016115.
[9] M. E. J. Newman and M. Girvan. "Finding and evaluating community structure in networks". Physical Review E, vol. 69, Feb. 2004, pp. 026113, DOI: 10.1103/PhysRevE.69.026113.
[10] M. E. J. Newman. "Fast Algorithm for Detecting Community Structure in Networks". Physical Review E, vol. 69, Jun. 2004, pp. 066133, DOI: 10.1103/PhysRevE.69.066133.
[11] A. Gog, D. Dumitrescu and B. Hirsbrunner. "Community Detection in Complex Networks Using Collaborative Evolutionary Algorithms". 9th European Conference on Artificial Life (ECAL 07), Springer Press, Sept. 2007, pp. 886-894, DOI: 10.1007/978-3-540-74913-4_89.
[12] X. Liu, D. Li, S. Wang and Z. Tao. "Effective Algorithm for Detecting Community Structure in Complex Networks Based on GA and Clustering". International Conference on Computational Science (ICCS 07). Springer Press, Jul. 2007, pp. 657-664, DOI: 10.1007/978-3-540-72586-2_95.
[13] C. Pizzuti. "GA-Net: A Genetic Algorithm for Community Detection in Social Networks". 10th International Conference on Parallel Problem Solving From Nature (PPSN 08), Springer Press, Sept. 2008, pp. 1081-1090, DOI: 10.1145/1569901.1570068.
[14] A. L. N. Fred and A. K. Jain. "Combining multiple clusterings using evidence accumulation". IEEE Transactions on Pattern Analysis and Machine Intelligence, vol. 27, Jun. 2005, pp. 835-850, 10.1109/TPAMI.2005.113.
[15] B. Yang, W. K. Cheung and J. Liu. "Community Mining from Signed Social Networks". IEEE Transactions on Knowledge and Data Engineering, vol. 19, Sept. 2007, pp. 1333-1348, DOI: 10.1109/34.868688.
[16] S. Fortunato and M. Barthélemy. "Resolution limit in community detection". Proc. Natl. Acad. Sci. USA, vol. 104, Jan. 2007, pp. 36-41, 10.1073/pnas.0605965104.
[17] L. I. Kuncheva, M. Skurichina, R. P. W. Duin. "An Experimental Study on Diversity for Bagging and Boosting with Linear Classifiers". Information Fusion, vol. 3, Dec. 2002, pp. 245-258, DOI: 10.1016/S1566-2535(02)00093-3.
[18] Y. Hong and S. Kwong. "To combine steady-state genetic algorithm and ensemble learning for data clustering". Pattern Recognition Letters, vol. 29, Jul. 2008, pp. 1416-1423, DOI: 10.1016/j.patrec.2008.02.017.
[19] W. W. Zachary. "An information flow model for conflict and fission in small groups". Journal of Anthropological Research, vol. 33, Apr. 1977, pp. 452-473.